\documentclass[twocolumn%
              ,superscriptaddress%
              ,aps%
              ,pra]{revtex4}
\usepackage{graphicx}
\usepackage{amssymb}
\usepackage{amsmath}
\usepackage{hyperref}
\usepackage{tikz}
\usepackage[utf8]{inputenc}

\def\cF{\mathcal{F}}
\def\cL{\mathcal{L}}

\def\cS{\mathcal{S}}

\def\kf{f}

\def\mB{\mathrm{B}}
\def\mD{\mathrm{D}}

\def\mmedium{\mathrm{m}}
\def\msphere{\mathrm{s}}
\def\meff{\mathrm{eff}}

\def\mfit{\mathrm{rm}}
\def\msc{\mathrm{sc}}
\def\mDvD{\mathrm{DvD}}
\def\mded{\mathrm{ded}}
\def\mDv{\mathrm{Dv}}
\def\mde{\mathrm{de}}

\begin{document}

\title{Universal Casimir interactions in the sphere-sphere geometry}

\author{Tanja Schoger}
\affiliation{Universität Augsburg, Institut für Physik, 86135 Augsburg, Germany}

\author{Benjamin Spreng}
\affiliation{Department of Electrical and Computer Engineering, 
University of California, Davis, CA 95616, USA}

\author{Gert-Ludwig Ingold}
\affiliation{Universität Augsburg, Institut für Physik, 86135 Augsburg, Germany}

\author{Astrid Lambrecht}
\affiliation{Forschungszentrum Jülich, 52425 Jülich, and 
RWTH Aachen University, 52062 Aachen, Germany}

\author{Paulo A. Maia Neto}
\affiliation{Instituto de F\'{\i}sica, Universidade Federal do Rio de Janeiro, Caixa Postal 68528,  Rio de Janeiro,  RJ, 21941-972, Brazil}

\author{Serge Reynaud}
\affiliation{Laboratoire Kastler Brossel, Sorbonne Université, CNRS, ENS-PSL, Collège de France,
Campus Jussieu, 75252 Paris, France}

\begin{abstract}
We study universal Casimir interactions in two configurations which appear as dual to each other. The first involves spheres described by the Drude model and separated by vacuum while the second involves dielectric spheres immersed in a salted solution at distances larger than the Debye screening length. In both cases, the long-distance limit, equivalently the high-temperature limit, is dominated by the effect of low-frequency transverse magnetic thermal fluctuations. They are independent of the details of dielectric functions of materials, due to the finite conductivity of metals in the former case and of salted water in the latter one. They also show universality properties in their dependence on geometric dimensions, in relation to an approximate conformal invariance of the reduced free energy.
\keywords{Casimir interaction, Screening in electrolytes, Drude conductivity.}
\end{abstract}

\maketitle

\section{Introduction}	

We review recent progress in the understanding of Casimir interactions \cite{Casimir1948,Lamoreaux1999,Milton2011,Decca2011,Gong2021} between spheres of arbitrary radii and focus our attention on two physical configurations which show analogous universality properties with interesting differences however. 

The first configuration corresponds to two metallic spheres described by a Drude conductivity model and separated by vacuum. This case has been shown to lead to a universal expression in the high-temperature limit or, equivalently, the large-distance limit, with the interaction not depending on the details of the electromagnetic response of the involved material \cite{CanaguierDurand2012,Bimonte2012}.
The universal thermal Casimir contribution \cite{Bostrom2000} dominates the nonuniversal terms at distances larger than the thermal wavelength $\lambda_T = \hbar c/k_\mB T$, of the order of $8\,\mu$m at room temperature. The weakness of the Casimir force in this regime makes its experimental detection challenging \cite{Sushkov2011}.

The second configuration corresponds to two spherical dielectric particles immersed in a conducting electrolyte solution, at the limit where electrostatic interactions are efficiently screened \cite{MaiaNeto2019}. This case also leads to a universal expression in the limit of high temperatures, with the free energy not depending on the detailed dielectric function of the involved material \cite{Schoger2021universal}. The universal thermal contribution now dominates the nonuniversal terms at much smaller distances of the order of $0.1\,\mu$m (this point will be made more precise in the following). The existence and magnitude of this universal Casimir interaction has recently been confirmed by measurements \cite{Pires2021} involving a silica microsphere held by optical tweezers in the vicinity of a larger sphere, both spheres being immersed in salted water.

The high-temperature Casimir interaction between spheres can be derived exactly in the two configurations described above by using the scattering approach \cite{Lambrecht2006,MaiaNeto2008}. We present below the expressions obtained in this manner for the two cases and discuss their analogies as well as their differences. We emphasize the role played by an approximate conformal invariance which highlights the connections between the electromagnetic Casimir effect and the critical Casimir effect \cite{Fisher1978,Hanke1998,Hertlein2008,Magazzu2019}, another long-range force appearing when fluctuations are confined within walls \cite{Parsegian2006,Gambassi2009}.

\section{Scattering formula for Casimir interaction}
\label{sec:Scatteringformula}

We consider the Casimir free energy for a setup consisting of two spheres embedded in a medium \cite{RodriguezLopez2011,Teo2012,Bimonte2018a,Bimonte2018b}. 
The spheres are placed at a center-to-center distance 
\begin{equation}
\mathcal{L} = L + R_1 + R_2 
\end{equation}
along the $z$-axis from each other, where $L$ is the smallest distance between the two spheres and $R_1$ and $R_2$ their radii.
In this section, we present the derivation of the high-temperature Casimir free energy, emphasizing the analogies between the two studied configurations and writing the associated scattering matrix elements. 

\subsection{High-temperature limit}

Within the scattering approach  \cite{CanaguierDurand2010}, the Casimir free energy at a temperature $T$ is given as a sum over the imaginary Matsubara frequencies \cite{Schwinger1978,Guerout2014}
\begin{equation}
\label{eq:free_energy}
\mathcal{F} = \frac{k_\mB T}{2}\sum_{n=-\infty}^\infty 
\mathrm{tr}\log\left(1 - \mathcal{M}(|\xi_n|)\right)
\,,\quad \xi_n = \frac{2\pi n k_\mB T}\hbar\,.
\end{equation}
The round-trip operator 
\begin{equation}
\mathcal{M} = \mathcal{R}_2 \mathcal{T}_{21} \mathcal{R}_1 \mathcal{T}_{12}
\end{equation}
accounts for the scattering process between the two spheres where $\mathcal{R}_{j}$ denotes
the reflection operator for an electromagnetic wave at sphere $j=1,\,2$ and $\mathcal{T}_{ij}$
represents the translation from the center of sphere $j$ to the center of sphere $i$. 

For the case of Drude spheres in vacuum, the high-temperature limit corresponds to distances $\mathcal{L}$ much larger than the thermal wave length $\lambda_T$. For the case of dielectric spheres in salted water, the discussion of this limit is more subtle as it involves details of the dielectric functions to be discussed at a later stage in this paper. In both cases, we will call high-temperature limit the case where the zero-frequency term $n=0$ is the dominant contribution to the Casimir free energy~\eqref{eq:free_energy}. The non-zero Matsubara frequencies lead  to exponentially small contributions at the high-temperature limit, due to the propagation factor appearing in translation operators $\mathcal{T}_{ij}$ (more precise discussion below). 

In this limit, the Casimir free energy 
\begin{equation}
\label{eq:free_energy_ht}
\mathcal{F}_T = -k_\mB T \,f
\end{equation}
has a universal form linear in the temperature and independent of the Planck constant. 
Here, the reduced free energy $f$ only depends on the geometric parameters and possibly on the electromagnetic response of the spheres and of the medium. The Casimir entropy $\cS = k_\mB  f$ is thus temperature independent, pointing towards the entropic origin of the Casimir free energy $\cF_T=-T\cS$.

After expanding the logarithm in Eq. \eqref{eq:free_energy} in a Mercator series, we find for the reduced free energy $f$
\begin{equation}
\label{eq:def_f}
f =  \sum_{r=1}^\infty f^{(r)}\,,
\quad
f^{(r)} = \frac{\mathrm{tr}\mathcal{M}^r(0)}{2r} \,,
\end{equation}
where the sum over $r$ accounts for the contributions $f^{(r)}$ to the free energy corresponding to $r$ round-trips.
As we restrict ourselves to the zero Matsubara frequency $\xi=0$ in the following, we will omit the corresponding argument of the round-trip operator $\mathcal{M}$ from now on.

For the evaluation of the trace in \eqref{eq:def_f}, we adopt the plane-wave basis \cite{Spreng2020,Schoger2021}. In view of the geometry of our problem, we decompose the wave vector
into a transverse part $\mathbf{k}$ with modulus $k$ and a $z$-component which after Wick rotation is written in terms of 
${\kappa = [\varepsilon_\mmedium(\xi/c)^2+\mathbf{k}^2]^{1/2}}$.
For $\xi=0$, this relation simplifies to $\kappa = k$. As by definition $k$ is positive,
we also need to specify the direction of propagation along the positive ($+$) or negative ($-$) $z$-direction.
Finally, the polarization $p = \mathrm{TM},\,\mathrm{TE}$ is taken with respect to the Fresnel plane spanned by the $z$-axis and the incident wave vector. For the $\mathrm{TM}$ modes, the electric field component lies within the Fresnel plane.
In both configurations studied in this paper, the Casimir free energy is due to TM modes while the contribution of TE modes vanishes.
For $\xi\to0$, the free energy contribution of $r$ round-trips 
\eqref{eq:def_f} is then given by \cite{Spreng2018}
\begin{equation}
\begin{aligned}
\label{eq:def_fr}
f^{(r)} &= \frac{1}{2r} \int\frac{\mathrm{d}\mathbf{k}_1\ldots 
	\mathrm{d}\mathbf{k}_{2r}}{(2\pi)^{4r}} 
\prod_{j=1}^{r} e^{-k_{2j}\mathcal{L}} e^{-k_{2j-1}\mathcal{L}}\\
&\qquad\quad\times \langle \mathbf{k}_{2j+1}, \text{TM}, - \vert
\mathcal{R}_2 \vert \mathbf{k}_{2j}, \text{TM}, + \rangle\\
&\qquad\quad\times\langle \mathbf{k}_{2j}, \text{TM}, + \vert
\mathcal{R}_1 \vert \mathbf{k}_{2j-1}, \text{TM}, - \rangle\,.
\end{aligned}
\end{equation}
In order to properly reproduce the trace, the indices are cyclic with $\mathbf{k}_{2r+1} = \mathbf{k}_1$. 
Without loss of generality, we moreover assumed that sphere 2 is above sphere 1 along the z-direction. 
The translation operator, which is diagonal in the plane-wave basis, is thus defined as
\begin{equation}
   \langle \mathbf{k}_{j}, \mathrm{TM}, +\vert
\mathcal{T}_{21} \vert \mathbf{k}_{i}, \mathrm{TM}, +\rangle 
= e^{-k_i \mathcal{L}} \delta(\mathbf{k}_j - \mathbf{k}_i)\,.
\end{equation}
The matrix elements of $\mathcal{T}_{12}$ are the same.

The TM-TM matrix elements of the reflection operators can be expressed in terms of the Mie scattering amplitudes \cite{BohrenHuffman2004}. Following the reasoning explained in detail in Appendix~B of Ref.~\onlinecite{Spreng2018} and adapting it to our specific configurations, we obtain their expressions in the limit $\xi\to0$
\begin{equation}
\label{eq:reflection_tm}
\langle \mathbf{k}_{j}, \mathrm{TM}\vert
\mathcal{R} \vert \mathbf{k}_{i}, \mathrm{TM}\rangle
= \frac{2\pi R}{k_j}\sum_{\ell = 1}^\infty 
 \mathcal{A}_\ell
 \frac{\chi^{2\ell}}{(2\ell)!}\,,
\end{equation}
with 
\begin{equation}
\label{eq:def_Aell}
\mathcal{A}_\ell=
\frac{\ell\big(\varepsilon_\msphere(0) - \varepsilon_\mmedium (0)\big)}
{\ell\varepsilon_\msphere(0) + (\ell+1)\varepsilon_\mmedium (0)}
\end{equation}
and
\begin{equation}
\label{eq:def_x}
\chi = 2 R \sqrt{k_i k_j}\cos\left(\frac{\varphi_{i} -\varphi_{j}}{2}\right) \,.
\end{equation}
The dielectric functions $\varepsilon_\msphere$ and $\varepsilon_\mmedium$ describe the non-magnetic materials constituting spheres and medium, respectively.
$\chi$ depends on the sphere radius $R$, the moduli of the in- and outgoing transverse wave vector $k_i$ and $k_j$, as well as the angle $\varphi_{i}-\varphi_{j}$ between the two wave vectors. In (\ref{eq:reflection_tm}), it is implicitly understood that the direction of propagation changes its sign upon reflection on each sphere.

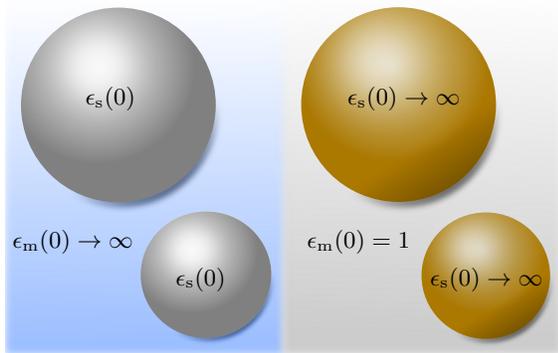
\begin{figure}
\centering
\begin{tikzpicture}
\node (img2) at (0,0) {\includegraphics[scale=0.37]{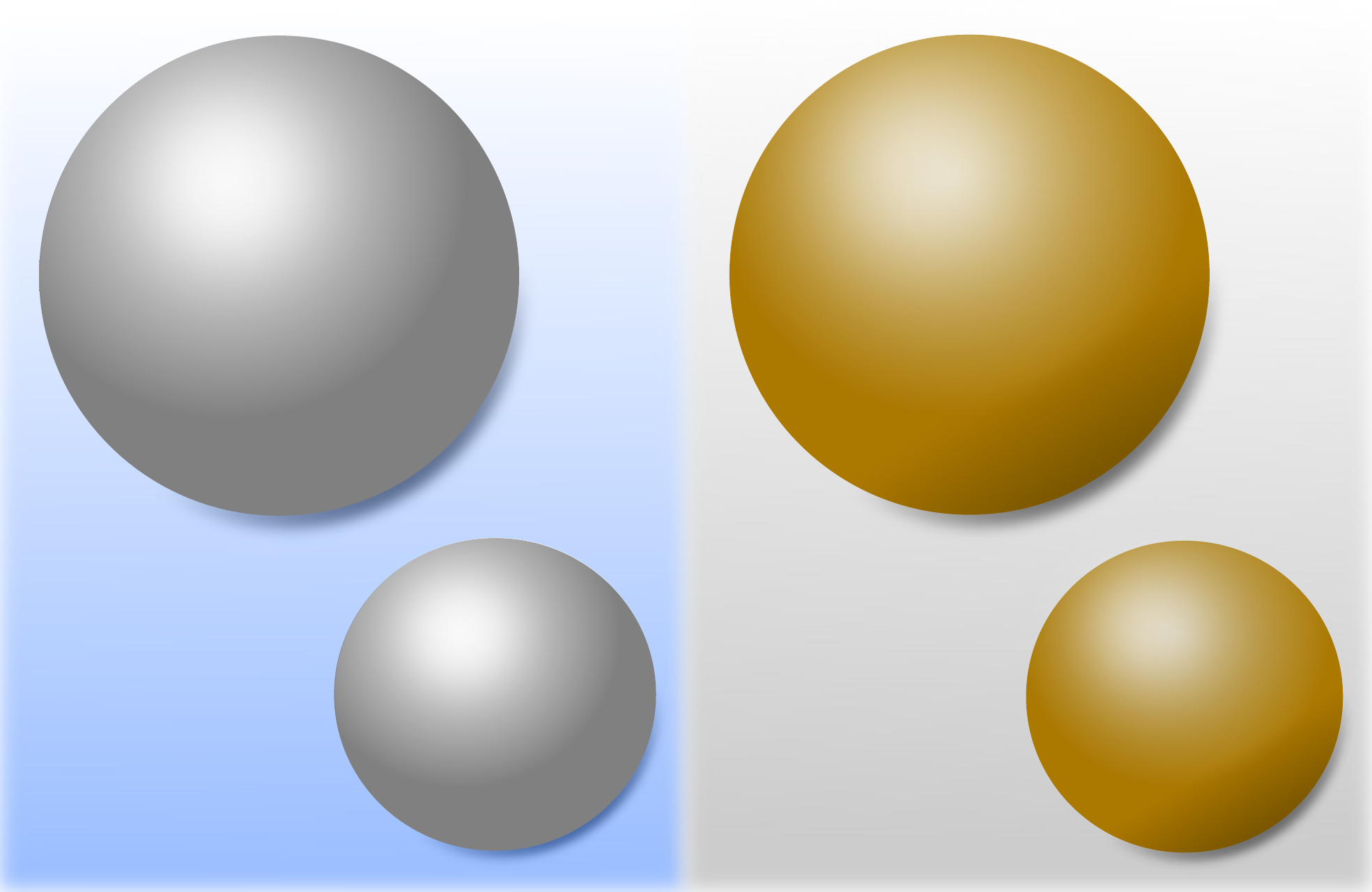}};
\node (A) at (-2.3,1) {$\epsilon_\msphere(0)$};
\node (B) at (1.6,1) {$\epsilon_\msphere(0) \to \infty$};
\node (C) at (-2.8,-0.9) {$\epsilon_\mmedium(0)\to \infty$};
\node (D) at (1,-0.9) {$\epsilon_\mmedium(0) =1$};
\node (E) at (-1.1, -1.4) {$\epsilon_\msphere(0)$};
\node (F) at (2.7, -1.4) {$\epsilon_\msphere(0) \to \infty$};
\end{tikzpicture}
\caption{Schematic representation of the two models studied in this paper, each model considering two spheres with the same properties immersed in a common medium. The figure on the left depicts two dielectric spheres with a finite dielectric constant $\epsilon_\msphere(0)$ in an electrolyte with finite conductivity so that $\epsilon_\mmedium(0)~\to~\infty$. 
For the case shown on the right, two metallic spheres have a finite conductivity and divergent dielectric function $\epsilon_\msphere(0)~\to~\infty$ in the static limit. The medium dielectric constant, which should be finite, is taken as that for vacuum $\epsilon_\mmedium(0)=1$.}
\label{fig:FigSetup}
\end{figure}

For the two cases of interest here, the model parameter $\mathcal{A}_\ell$ is independent of the details of the material properties. Metals with a finite conductivity can be described by the Drude model for which the dielectric function $\varepsilon_\msphere$ diverges for $\xi\to 0$ (see Fig.~\ref{fig:FigSetup}, right panel). The parameter $\mathcal{A}_\ell$ is thus independent of the details of the response function for the medium as long as $\varepsilon_\mmedium$ remains finite in that limit. An important example is the case with two Drude spheres separated by vacuum ($\mDvD$ model). For dielectrics immersed in a conducting electrolyte ($\mded$ model), the reverse situation occurs with the finite conductivity of the medium leading to a divergence of  $\varepsilon_\mmedium$ for $\xi\to0$ (see Fig.~\ref{fig:FigSetup}, left panel). Again, $\mathcal{A}_\ell$ is independent of material properties of the spheres provided $\varepsilon_\msphere$ remains finite. The parameters for the two models are summarized in Table~\ref{model_param}. The different expressions for $\mathcal{A}_\ell$ are deduced from the ratios $\varepsilon_\msphere(0)/\varepsilon_\mmedium(0)$, which are infinite and null respectively for the two models.

\begin{table}[t]
 \centering
   \caption{Parameter $\mathcal{A}_\ell$ determining the reflection matrix element (\ref{eq:reflection_tm}) for TM modes for the two models studied in the paper. }
{\begin{tabular}{c|c|c}
sphere--medium--sphere model & $\mathcal{A}_\ell$ & $~\dfrac{\varepsilon_\msphere(0)}{\varepsilon_\mmedium(0)}$~ \\
\hline
Drude--vacuum--Drude ($\mDvD$)& 1 & $\infty$ \\
\hline
dielectric--electrolyte--dielectric ($\mded$)& $~-\dfrac{\ell}{\ell+1}$ ~& 0
\end{tabular}
\label{model_param}}
\end{table}

\subsection{Free-energy contributions for $r$ round-trips for the two models}

We now consider the two models listed in Table~\ref{model_param} and depicted in Fig.~\ref{fig:FigSetup}, where the Casimir interaction becomes universal, \textit{i.e.} independent of the detailed material properties. General expressions for the round-trip expansion are derived for the two cases in parallel. The evaluation of the integrals will then be carried out in Section~\ref{sec:Results} and a qualitative discussion of the results presented in Section~\ref{sec:Discussion}. 

For both models, only the reflection matrix elements for TM modes contribute and the corresponding sum \eqref{eq:reflection_tm} can be evaluated analytically. For a Drude sphere in vacuum, denoted by the superscript $\mDv$, we find 
\begin{equation}
\label{eq:r_dr}
\langle \mathbf{k}_{j}, \mathrm{TM}\vert
\mathcal{R}^{(\mDv)} \vert \mathbf{k}_{i}, \mathrm{TM}\rangle
= \frac{2\pi R}{k_j} \left[
\cosh\left(\chi\right)
- 1
\right] \,,
\end{equation}
while for a dielectric sphere in an electrolyte (superscript $\mde$), the sum \eqref{eq:reflection_tm} yields 
\begin{multline}
\label{eq:r_die_explicit}
\langle \mathbf{k}_{j}, \mathrm{TM}\vert
\mathcal{R}^{(\mde)} \vert \mathbf{k}_{i}, \mathrm{TM}\rangle\\
= -\frac{2\pi R}{k_j} \left[\cosh(\chi) 
+ 2\frac{\cosh(\chi)-1}{\chi^2} - 
2\frac{\sinh(\chi)}{\chi}
\right]\,.
\end{multline}
For the evaluation of the integrals in (\ref{eq:def_fr}), we found it more convenient to employ the equivalent integral representation 
\begin{multline}
 \label{eq:r_die}
 \langle \mathbf{k}_{j}, \mathrm{TM}\vert
	\mathcal{R}^{(\mde)} \vert \mathbf{k}_{i}, \mathrm{TM}\rangle\\
= -\frac{2\pi R}{k_j} \int_0^1  \mathrm{d}t
 \left[\cosh(\chi) - 2 t \cosh(t\chi)
\right] 
\end{multline}
instead of the expression (\ref{eq:r_die_explicit}) which contains inverse powers of $\chi$. 
The coupling of the angles $\varphi_i$ and $\varphi_j$ of incident and reflected waves, respectively, through the cosine in (\ref{eq:def_x}) can be avoided by introducing Cartesian coordinates 
\begin{equation}
x_i = (k_i\mathcal{L})^{1/2}\cos(\varphi_i/2)\,,\quad
y_i = (k_i\mathcal{L})^{1/2}\sin(\varphi_i/2)\,.
\end{equation}

For two Drude spheres in vacuum, the contribution (\ref{eq:def_fr}) from $r$ round-trips to the dimensionless free energy $\kf$ then becomes
\begin{widetext}
\begin{equation}
\label{eq:fr_dr}
f^{(r)}_\mDvD = \frac{1}{2r} \left(\frac{R_1 R_2}{\pi^{2}\mathcal{L}^2}\right)^r
\int_{-\infty}^\infty \mathrm{d} \mathbf{x} 
\int_{-\infty}^\infty \mathrm{d} \mathbf{y}  
\prod_{j=1}^r 
e^{-\left(x_{2j}^2 + y_{2j}^2\right)} 
e^{-\left(x_{2j-1}^2 + y_{2j-1}^2\right)}
\left[\cosh(\chi_{2j}^{(2)}) - 1\right] 
\left[\cosh(\chi_{2j-1}^{(1)}) - 1\right] \,,
\end{equation}
whereas for two dielectric spheres in an electrolyte, we get
\begin{equation}
\begin{aligned}
\label{eq:fr_die}
f^{(r)}_\mded &= \frac{1}{2r} \left(\frac{R_1 R_2}{\pi^{2}\mathcal{L}^2}\right)^r
\int_0^1 \mathrm{d} \mathbf{t}
\int_{-\infty}^\infty \mathrm{d} \mathbf{x} 
\int_{-\infty}^\infty \mathrm{d} \mathbf{y}  
\prod_{j=1}^r 
e^{-\left(x_{2j}^2 + y_{2j}^2\right)} 
e^{-\left(x_{2j-1}^2 + y_{2j-1}^2\right)}
\\
&\qquad\qquad\qquad\times 
\left[\cosh(\chi_{2j}^{(2)}) -
 2t_{2j} \cosh(t_{2j}\chi_{2j}^{(2)})\right]
\left[\cosh(\chi_{2j-1}^{(1)}) - 
2t_{2j-1}\cosh(t_{2j-1}\chi_{2j-1}^{(1)})
\right]\,.
\end{aligned}
\end{equation}
\end{widetext}
The argument \eqref{eq:def_x} of the hyperbolic cosine 
in Cartesian coordinates is given by
\begin{equation}
\label{eq:chi_cartesian}
\chi_i^{(n)} = \frac{2R_n}{\cL}(x_ix_{i+1} + y_i y_{i+1})\,.
\end{equation}
The subscript $i$ denotes the $i$-th reflection during the $r$ round-trips
and the superscript $(n)$ refers to the sphere at which the reflection
occurs. The integrals over $\mathbf{x},\, \mathbf{y}$ and $\mathbf{t}$ are of dimension $2r$ each.

Expanding the product in both expressions, we find that they have
one term in common which consists of products of all $\cosh(\chi_j^{(n)})$ factors and is given by
\begin{widetext}
\begin{equation}
\label{eq:def_fr_sc}
f_\msc^{(r)} = \frac{1}{2r}\left(\frac{R_1 R_2}{\pi^{2}\mathcal{L}^2}\right)^r
\int_{-\infty}^\infty \mathrm{d} \mathbf{x} 
\int_{-\infty}^\infty \mathrm{d} \mathbf{y}  
\prod_{j=1}^r 
e^{-\left(x_{2j}^2 + y_{2j}^2\right)} 
e^{-\left(x_{2j-1}^2 + y_{2j-1}^2\right)}
\cosh(\chi_{2j}^{(2)})\cosh(\chi_{2j-1}^{(1)})\,.
\end{equation}
\end{widetext}
This term will be discussed below as the scalar result denoted by a subscript sc.

\section{Universal Casimir interaction and conformal invariance}
\label{sec:Results}

In this section, we present results corresponding to the two studied configurations and compare them to other known expressions. 

\subsection{Scalar field with Dirichlet boundary conditions}

We begin with the model corresponding to spheres with Dirichlet boundary conditions for a scalar field. The Casimir free energy for this formal model has been calculated in many publications,
for example in Ref. \onlinecite{Eisenriegler1995} and \onlinecite{Zhao2013}, where the configuration of two exterior spheres was mapped to two concentric spheres by using conformal invariance (details given below). Bispherical coordinates were used in Ref. \onlinecite{Bimonte2012} while the spherical- and plane-wave basis were used in Ref. \onlinecite{Teo2014} and Ref. \onlinecite{Schoger2021} respectively to derive the Casimir free energy in the scalar case. 

This scalar model corresponds also to the expression \eqref{eq:def_fr_sc}, found as a part of more complete expressions for our two electrodynamical calculations. After expansion of the hyperbolic cosines into exponential functions, this expression is transformed into a sum of Gaussian integrals where the bilinear form in the exponent can be expressed in terms of symmetric circulant tridiagonal matrices. Such matrices can be given an equivalent meaning in terms of a tight-binding model on a ring with the diagonal matrix elements corresponding to the local potential on the lattice sites and the non-diagonal matrix elements describing the hopping between the sites. Their determinants can be expressed in terms of transfer matrices \cite{Molinari1997,Molinari2008}, leading to the analytical expression
\begin{equation}
\label{eq:fr_sc}
f_\msc^{(r)} = 
\frac{1}{4r}\frac{\cosh(r\varpi)}{\sinh^2(r\varpi)}
\,,\quad \varpi = \mathrm{arcosh}(y) \,,
\end{equation}
where
\begin{equation}
\label{eq:def_y}
y =\frac{\mathcal{L}^2 - R_1^2 - R_2^2 }{2R_1R_2}
= 1 + \frac{L}{R_\meff} + \frac{u}{2} \left(\frac{L}{R_\meff}\right)^2\,.
\end{equation}
Apart from the surface-to-surface distance $L$, the parameter $y$ depends on the radii through the effective radius 
\begin{equation}
\label{eq:def_reff}
R_\mathrm{eff} = \frac{R_1 R_2}{R_1 + R_2}
\end{equation}
and the parameter $u$ measuring the ratio of radii in a symmetric manner
\begin{equation}
\label{eq:def_u}
u = \frac{R_1 R_2}{(R_1+R_2)^2}
\quad\Leftrightarrow\quad
\frac1u = \frac{R_1}{R_2}+\frac{R_2}{R_1}+2\,.
\end{equation}
The parameter $u$ ranges from $u=0$ for the plane-sphere geometry to $u=1/4$ for two spheres of equal radii.

The fact that the high-temperature Casimir free energy for a scalar field is a function of the parameter $y$ alone means that it obeys exact conformal invariance, a property which makes this problem analogous to the critical Casimir effect \cite{Burkhardt1995}. 
The conformal invariance of $y$ which allows one to simplify calculations in a two-spheres geometry has been known for a long time. Thomson used this property to calculate the electrostatic capacitance between two spherical conductors \cite{Thomson1847,Thomson1853}. The dimensionless part of the mutual capacitance $C_{12}$ is for example written as a function of $\varpi = \mathrm{arcosh}(y)$ in \S173 of Maxwell's Treatise \cite{Maxwell1873}
\begin{equation}
\label{eq:C12}
C_{12}= - \frac{4\pi\varepsilon_0 R_1R_2}{\cL} 
\sum_{n=0}^\infty \frac{\sinh(\varpi)}{\sinh((n+1)\varpi)} \,.
\end{equation}

The mathematical meaning of this invariance has also been known for a long time \cite{Liouville1847,Bromwich1900,Klein1926}. The group of conformal transformations in 3D Euclidean space is generated by isometries and inversion with respect to an arbitrary center. Spheres are transformed into spheres by this group if one includes planes as particular spheres with center at infinity. The quantity generalizing the squared Euclidean distance to this geometry of spheres is $\cL^2-R_1^2-R_2^2$ with $\cL$ the distance of their centers and $R_1,R_2$ their radii. Under an arbitrary conformal transformation, this quantity is simply multiplied by the product $\lambda_1\lambda_2$ of two conformal factors associated with each sphere. As these factors can also be understood as the multiplicative factors appearing in the transformation of the radius of each sphere, it follows that the quantity $y$ defined in \eqref{eq:def_y} is preserved by the transformations. The case of spheres intersecting at right angle corresponds to $y=0$, which is preserved by an arbitrary conformal transformation. More generally, $y$ is a generalized angle of intersection of the two spheres (equal to the imaginary number $\imath\,\varpi$ for exterior spheres).

For the cases studied in this paper, Drude spheres in vacuum and dielectric spheres in a conducting electrolyte, the property of conformal invariance will be met approximately but not exactly. Precisely the reduced free energy will be largely determined by the invariant quantity $y$ but also affected by the parameter $u$. For this reason, it will be denoted $\kf_u(y)$, with the argument $y$ and subscript $u$ occasionally dropped when their presence is not needed. 

\subsection{Drude spheres in vacuum and dielectric spheres in electrolyte}

In contrast to the scalar case discussed before, we do not have general expressions for the reduced free energy \eqref{eq:fr_dr} and \eqref{eq:fr_die}
for an arbitrary number of round-trips. In the case of two Drude spheres however, an exact expression is known for the sum over the contributions of all round-trips \cite{Schoger2021}. 
The difference with respect to the scalar case consists in the absence of the monopolar term associated with $\ell=0$ in the reflection matrix element (\ref{eq:reflection_tm}) giving rise to the extra term $-1$ in (\ref{eq:r_dr}). The bilinear form appearing in the Gaussian integrals in (\ref{eq:fr_dr}) is then no longer given by a circulant matrix but by block-diagonal matrices. The evaluation of the Gaussian integrals can be mapped onto a combinatorial problem of finding bicolored integer partitions as discussed in detail in Ref.~\onlinecite{Schoger2021}. In the end, it turns out that the partitions are related to the elements
of the dimensionless capacitance matrix of two spherical conductors. 
The free energy for the transverse magnetic modes is then expressed in terms of the determinant of a dimensionless capacitance matrix \cite{Fosco2016} 
\begin{equation}
\label{eq:f_dr}
f_\mDvD = f_\msc - \frac{1}{2}\log\left( \mathrm{det}\,\mathbf{c}\right)\,,
\end{equation}
with the latter related to the full capacitance matrix through
\begin{equation}
\label{eq:capa_mat}
\mathbf{C} =  4\pi\varepsilon_0\sqrt{R_1R_2} \,\mathbf{c}
\,,\quad \mathbf{c} =
\left(
\begin{array}{cc}
c_{11}	&	c_{12}	 \\
c_{12}	&	c_{22}	 
\end{array}
\right) \,.
\end{equation} 
The dimensionless capacitance matrix coefficients are \cite{Thomson1853,Maxwell1873,Lekner2011,Smythe1988} 
\begin{equation}
\label{eq:Cij}
\begin{aligned}
c_{11} &= \sum_{n=0}^\infty \frac{\sqrt{R_1R_2} \sinh(\varpi)}{R_1 \sinh(n\varpi) + R_2 \sinh((n+1)\varpi)} \,,\\
c_{22} &= \sum_{n=0}^\infty \frac{\sqrt{R_1R_2} \sinh(\varpi)}{R_2 \sinh(n\varpi) + R_1 \sinh((n+1)\varpi)}\,,\\
c_{12} &=- \sum_{n=0}^\infty \frac{\sqrt{R_1R_2}\sinh(\varpi)}{\cL\sinh((n+1)\varpi)}\,.
\end{aligned}
\end{equation}
Reference \onlinecite{Schoger2021} derives the full expression of the free energy for two Drude spheres of arbitrary radii within the plane-wave basis, while Ref.~\onlinecite{Fosco2016} obtains a general expression for the difference between the full free energy of the electromagnetic field and the scalar field for arbitrary conductors by means of a path-integral approach to the free energy. 
Results of Ref.~\onlinecite{Schoger2021} agree with those presented for the plane-sphere geometry in Ref.~\onlinecite{Bimonte2012} and also with those for equally-sized spheres in Ref.~\onlinecite{Fosco2016}.

Next, we analyse the reduced free energy for two dielectric spheres in salted water \eqref{eq:fr_die}, 
which can also be written as
\begin{equation}
\begin{aligned}
f^{(r)}_\mded &= \frac{1}{2r} \left(\frac{4 R_1 R_2}{\pi^{2}\mathcal{L}^2}\right)^r
\int_0^1 \mathrm{d} \mathbf{t}
\prod_{i=1}^{2r} t_i \left[\delta(t_{i}-1) - 1\right]
\\&\times
\int_{-\infty}^\infty \mathrm{d} \mathbf{x} 
\int_{-\infty}^\infty \mathrm{d} \mathbf{y}  
\prod_{j=1}^r 
e^{-\left(x_{2j}^2 + y_{2j}^2\right)} 
e^{-\left(x_{2j-1}^2 + y_{2j-1}^2\right)}
\\&\times
\cosh(t_{2j}\chi_{2j}^{(2)}) 
\cosh(t_{2j-1}\chi_{2j-1}^{(1)}) \,,
\end{aligned}
\end{equation}
where the terms without a $t_i$-dependence are accounted for by introducing delta functions. The Gaussian integral can be evaluated as in the cases discussed above 
\begin{equation}
\label{eq:fr_die_matrix}
\begin{aligned}
f^{(r)}_\mded &=  \frac{1}{4r}\left(\frac{4R_1 R_2}{\mathcal{L}^2}\right)^r
\sum_{\sigma = \pm 1}
\int_0^1 \mathrm{d} \mathbf{t}
\frac{\prod_{i=1}^{2r} t_i \left[\delta(t_{i}-1) - 1\right]}
{\mathrm{det}\mathbf{M}^\sigma_{r}(\mathbf{t})}
\end{aligned}\,,
\end{equation}
where the $2r$-dimensional periodic tridiagonal matrix $\mathbf{M}^\sigma_{r}$ is given by
\begin{widetext}
\begin{equation}
\label{eq:def_M}
\mathbf{M}^\sigma_{r}(\mathbf{t}) = \left(
\begin{array}{cccccc}
1	&	t_1 R_1/\mathcal{L}	&	0	&	\ldots	&	0	&	\sigma t_{2r}R_2/\mathcal{L}  \\
t_1 R_1/\mathcal{L}	&	1	&	t_2 R_2/\mathcal{L}	&	& 	&	0 \\
0	&	t_2 R_2/\mathcal{L}	&	1	&	\ddots	&	&	\vdots \\
\vdots	&	&	\ddots	&	\ddots 	&	& 0  \\
0 & & & & & t_{2r-1} R_1/\mathcal{L} \\
\sigma  t_{2r} R_2/\mathcal{L}  & 0 & \ldots & 0 & t_{2r-1}R_1/\mathcal{L} & 1
\end{array}
\right)\,.
\end{equation}
\end{widetext}
As mentioned earlier, this matrix can be related to a tight-binding model on a ring. While for the scalar case, the hopping matrix element between adjacent sites varied periodically between two values, here the hopping matrix elements generically are nonperiodic. This difference illustrates the difficulty of finding closed analytical expressions for the reduced free energy $f_\mded$ for a given number of round-trips or even the sum over all round-trips. In the next section, we will present analytical results for a single round-trip (cf. Eq.~\eqref{eq:f1_die}) but resort to numerical results otherwise.

\begin{figure}[b]
\centering
\includegraphics[width=\columnwidth]{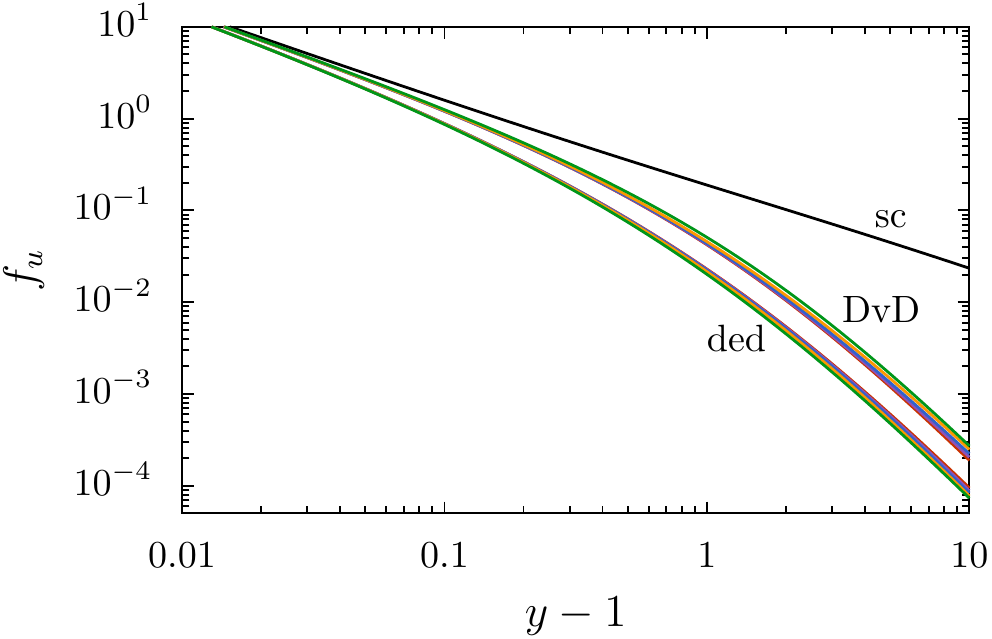}
\caption{Reduced free energy $f_u$ as function of the invariant geometric parameter $y-1$, at fixed $u$. The transition from the plane-sphere geometry to two spheres with equal radii is shown by $u = 0$ (red), $0.016$ (violet), $0.04$ (blue), $0.1$ (yellow) and $0.25$ (green). The upper curves correspond to Drude spheres in vacuum (DvD) and the lower ones to dielectric spheres in an electrolyte (ded). The two cases are compared to the scalar case (sc), where the reduced free energy is independent of $u$ (black).}
\label{fig:FigFuvsYm}
\end{figure}

We close this section by showing in Fig.~\ref{fig:FigFuvsYm} the reduced free energy $\kf_u$ as function of $y-1$ for the DvD and ded models (colored curves) and comparing it to the reduced free energy for the scalar model (black curve). The latter result is conformally invariant, as it depends on $y-1$ but not on $u$. For the DvD model (upper set of curves) and the ded model (lower set of curves), $\kf_u$ is mainly determined by $y-1$ while it also depends on the value of $u$. This variation is shown by the colors from the plane-sphere geometry, $u=0$ shown as red curve, to $0.016$ (violet), $0.04$ (blue), $0.1$ (yellow), and finally $u=0.25$ (green). This variation remains in a rather thin band between the red and the green curve, indicating that while the DvD and ded models are not exactly conformally invariant, the dependence on $u$ is weak and visible mainly when the distance between spheres becomes large, \textit{i.e.} for large values of $y$. 

\begin{figure}[b]
    \centering
    \includegraphics[width=\columnwidth]{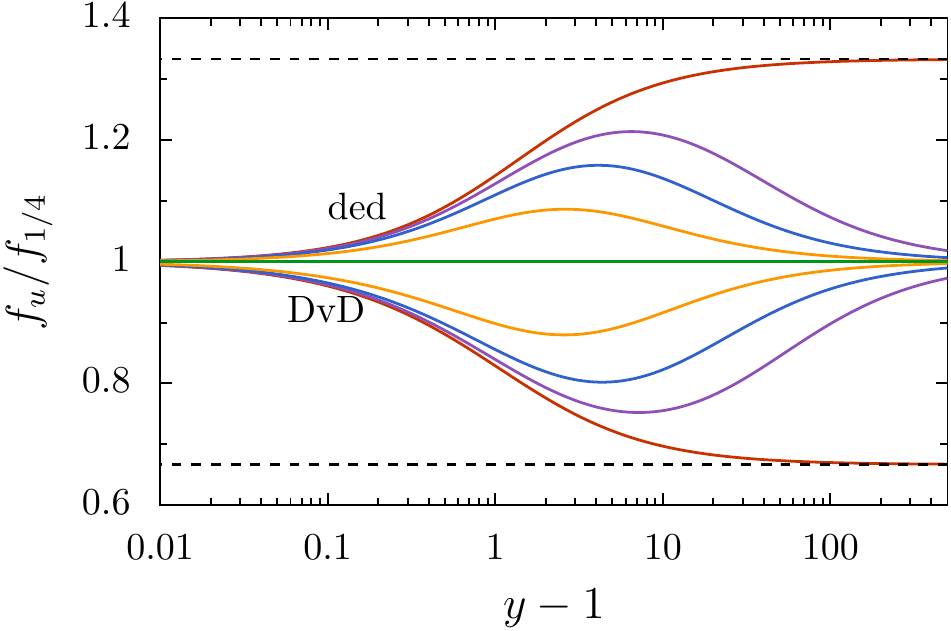}
    \caption{Ratio $f_u/f_{1/4}$ versus $y-1$ at fixed $u$, with the same conventions as in Fig.~\ref{fig:FigFuvsYm}. 
    The variations of the DvD curves (ratio below 1) and ded curves (ratio above 1) show opposite behavior. The red curves ($u=0$) have asymptotic values at 2/3 and 4/3 respectively. }
    \label{fig:ratioFuonF0vsYm} 
\end{figure}

We note that while the reduced free energy exhibits the same power law at $y\to 1$ for all models, this is no longer the case for large $y$, where the reduced free energy drops much faster for the electromagnetic cases than for the scalar case. Furthermore, the DvD and ded models have different variations, as is highlighted by drawing in Fig.~\ref{fig:ratioFuonF0vsYm} the ratio $f_u/f_{1/4}$ versus $y-1$ at fixed $u$. For the DvD and ded models respectively, this ratio is always below and above unity. It tends to unity at both limits of low or high values of $y-1$, except for the (red) curve $u=0$ corresponding to the plane-sphere configuration at large distances. These results will be given a simple qualitative interpretation in the following sections. 

\section{Limiting cases and approximate result for arbitrary distances}
\label{sec:Discussion}

In this section, we present new analytical and numerical results which allow one to better understand the analogies and the differences of the two configurations. 

\subsection{Single round-trip contribution}

While no explicit expression is known for an arbitrary fixed number $r$ of round-trips, we have been able to obtain analytical expressions \eqref{eq:fr_dr} and \eqref{eq:fr_die} for the DvD and ded cases, respectively, for the case of a single round-trip $r=1$. These results will be useful in the large-distance limit where this contribution is dominant but they will also lead to a good approximation at arbitrary distances between the spheres.

As already mentioned previously, the free energies for the DvD and ded models contain a contribution corresponding to a scalar field satisfying Dirichlet boundary conditions. From \eqref{eq:fr_sc} we obtain for the scalar contribution of a single round-trip
\begin{equation}
\label{eq:f1_sc}
 f^{(1)}_\msc = \frac{y}{4\left(y^2-1\right)}\,.
\end{equation}
This result is significant insofar as it represents the diverging contribution to the free energies of the DvD and ded models in the small distance limit $y\to1$. For later purposes, it is useful to note that for the scalar case, the contributions of arbitrary round-trips diverge for small distances. From \eqref{eq:fr_sc}, one finds $f^{(r)}_\msc = f^{(1)}_\msc/r^3$ for $y$ close to 1 or, equivalently, $\varpi$ close to 0.

We can then express the single round-trip contribution to the reduced free energy of two Drude spheres in vacuum as
\begin{widetext}
\begin{equation}
\label{eq:f1_DvD_integral}
f^{(1)}_\mDvD = 
f^{(1)}_\msc 
+ \frac{R_1 R_2}{2\pi^{2}\mathcal{L}^2}
\int_{-\infty}^\infty \mathrm{d} \mathbf{x} 
\int_{-\infty}^\infty \mathrm{d} \mathbf{y}  
e^{-\left(x_{2}^2 + y_{2}^2\right)} 
e^{-\left(x_{1}^2 + y_{1}^2\right)}
\left[1 -  \cosh(\chi_{1}^{(1)})- \cosh(\chi_{2}^{(2)})\right]\,,
\end{equation}
\end{widetext}
with $\chi_{i}^{(n)}$ defined in \eqref{eq:chi_cartesian}.
As in the scalar case, an expansion of the hyperbolic cosines in \eqref{eq:f1_DvD_integral} into exponentials
leads to Gaussian-type integrals which can easily be evaluated.
The Gaussian integrals yield expressions depending on 
the dimensionless radii $R/\mathcal{L}$. In view of our discussion of conformal invariance however, it is more convenient to express the results in terms of the parameter $y$ in Eq. \eqref{eq:def_y}. We thus introduce new notations representing the ratios of the two radii
\begin{equation}
\label{eq:def_alpha}
\alpha_{1,2} = \frac{R_{1,2}}{R_{2,1}} = \frac{1- 2u \pm \sqrt{1-4u}}{2u}   \,,\quad \alpha_{1} \alpha_{2} =1\,,
\end{equation}
with $\alpha_{1}$ corresponding to the sign $+$ in \eqref{eq:def_alpha} if $R_1>R_2$, and to the sign $-$ otherwise. These relations may also be expressed in terms of the geometric parameter $u$ defined in \eqref{eq:def_u} and
\begin{equation}
\label{eq:def_z}
 z = 2y + \alpha_1 + \alpha_2 
 = 2\left(y-1\right) + \frac1u = \frac{\cL^2}{R_1R_2}\,.
\end{equation}

We then find for the single round-trip contribution to the reduced free energy within the DvD model
\begin{equation}
\label{eq:f1_dr}
f^{(1)}_\mDvD = f^{(1)}_\msc + \frac{1}{2z} 
- \frac{1}{2}\sum_{n = 1,2}\frac{1}{2y + \alpha_n}\,.
\end{equation}
This expression is simplified for two equal-sized spheres ($u=0.25$)
\begin{equation}
\label{eq:f1_dr_u025}
f^{(1)}_{\mDvD, u= 0.25} = \frac{3}{4(2y+1)(y^2-1)}\,,
\end{equation}
and for the plane-sphere geometry $u=0$
\begin{equation}
\label{eq:f1_dr_u0}
f^{(1)}_{\mDvD,u=0} = \frac{1}{4y(y^2-1)}\,.
\end{equation}

We then consider the single round-trip expression for two dielectric spheres in an electrolyte. For $r=1$, the matrix \eqref{eq:def_M} takes the simple form
\begin{equation}
  \mathbf{M}^\sigma_{1}(\mathbf{t}) =
  \begin{pmatrix}
   1 & \dfrac{1}{\cL}(t_1R_1+\sigma t_2R_2)\\
   \dfrac{1}{\cL}(t_1R_1+\sigma t_2R_2) & 1
  \end{pmatrix} \,,
\end{equation}
and its determinant is obtained in terms of the quantities defined in \eqref{eq:def_alpha} and \eqref{eq:def_z}
\begin{equation}
\mathrm{det}\mathbf{M}^\sigma_{1}(\mathbf{t}) 
= 1- \frac{1}{z}\left(\sqrt{\alpha_1}t_1  + \sigma\sqrt{\alpha_2} t_2 \right)^2 \,.
\end{equation}
The reduced free energy \eqref{eq:fr_die} for a single round-trip can thus be expressed as
\begin{widetext}
\begin{equation}
f^{(1)}_\mded =  f^{(1)}_\msc 
+  \sum_{\sigma = \pm 1}
\int_0^1  \mathrm{d} \mathbf{t}
\frac{t_1 t_2}{z- \left(\sqrt{\alpha_1}t_1  + \sigma\sqrt{\alpha_2} t_2 \right)^2}
\left[1-\delta(t_1-1)-\delta(t_2-1)\right] \,,
\end{equation}
\end{widetext}
where we again isolated the result \eqref{eq:f1_sc} for a scalar field. Evaluation of the integrals over $t_1$ and $t_2$ finally yields
\begin{equation}
\label{eq:f1_die}
\begin{aligned}
f^{(1)}_\mded &= f^{(1)}_\msc
 + \frac{z}{12}\log \left[\frac{z^2(y^2-1)}{(yz + 1/2)^2} \right]
 \\&+\frac{1}{12\sqrt{z}}\sum_{n = 1,2}\frac{1}{\alpha_n^{3/2}}
 \log \left[
 \frac{2y^2 + \alpha_n y -1 +\sqrt{\alpha_n z}}
 {2y^2 + \alpha_n y -1 -\sqrt{\alpha_n z}}
 \right]\,.
 \end{aligned}
\end{equation}
This expression is read for two spheres with the same radius ($u=0,25$)
\begin{equation}
\begin{aligned}
\label{eq:f1_die_u025}
f_{\mded,u=0.25}^{(1)} &=
\frac{y}{4(y^2-1)}
+ \frac{y+1}{6} \log\left[\frac{(y^2-1)(y+1)^2}{(y+1/2)^4}\right] \\
&+ \frac{1}{6\sqrt{2(y+1)}}\log\left[\frac{2y-1+\sqrt{2}/\sqrt{y+1}}%
{2y-1-\sqrt{2}/\sqrt{y+1}}\right]\,,
\end{aligned}
\end{equation}
while it gets a simpler form for the special plane-sphere geometry ($u=0$)
\begin{equation}
\label{eq:f1_die_u0}
f_{\mded,u=0}^{(1)} =
\frac{y}{4}\left[\frac{1}{y^2-1}+\log\left(\frac{y^2-1}{y^2}\right)\right]\,.
\end{equation}

\begin{figure}[b]
\centering
\includegraphics[width=\columnwidth]{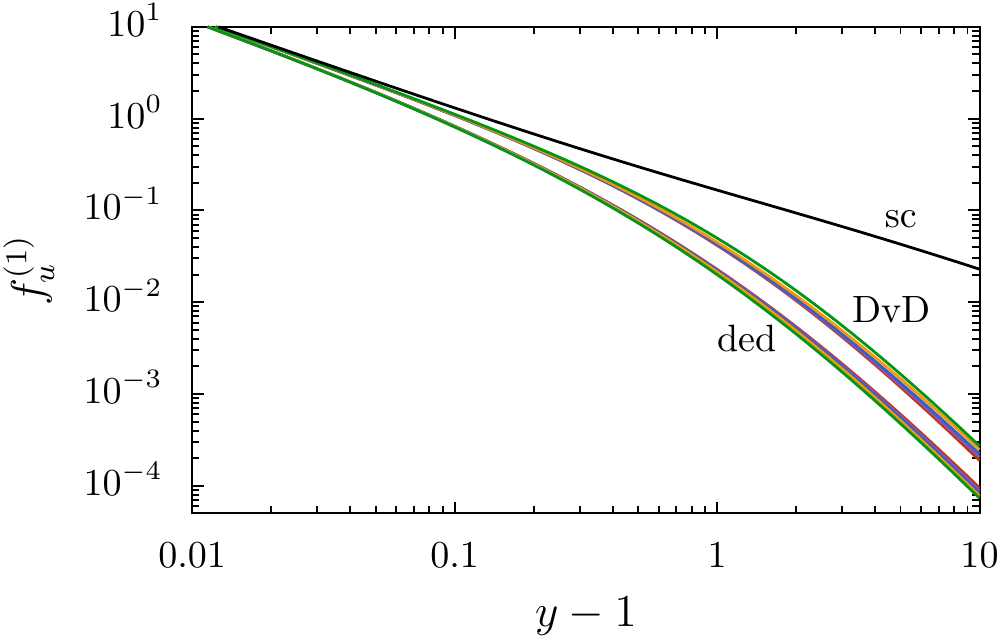}
\caption{Single round-trip contribution $f_u^{(1)}$ to the reduced free energy versus $y-1$, with the same conventions as in Fig.~\ref{fig:FigFuvsYm}. The upper curves correspond to Drude spheres in vacuum (Eq.~\ref{eq:f1_dr}) and the lower ones to dielectric spheres in an electrolyte (Eq.~\ref{eq:f1_die}). The two cases are compared to the scalar case (Eq.~\eqref{eq:f1_sc}), where the result is independent of $u$ (black). }
\label{fig:FigFu1vsYm}
\end{figure}

Fig.~\ref{fig:FigFu1vsYm} displays the dependence of the single round-trip contribution to the reduced free energy, as a function of $y-1$ for different values of $u$. The upper and lower curves represent the results for the DvD and ded models respectively. In each case, the red, violet, blue, yellow, and green curves correspond to $u=0, 0.016, 0.04, 0.1,$ and $0.25$. The scalar result, not depending on $u$, is shown as the black curve. One sees that all curves tend to be the same in the small-distance limit $y\to 1$, which means that they have the same divergence there. At large distances in contrast, the curves show different behaviors which will be discussed below by using the results of a multipolar expansion. We may also note at this point that the single round-trip contributions dominate the full expressions in the large-distance limit, so that the curves thus tend to be the same asymptotic values on Fig.~\ref{fig:FigFuvsYm} and  Fig.~\ref{fig:FigFu1vsYm}.

\subsection{Large and small distance limits}
We now discuss the limits of large and small distances, which will allow us to better understand the origin of breaking of conformal invariance in DvD and ded models. 

At large distances, it is sufficient to consider the dipolar contribution ($\ell =1$) to the reflection matrix elements \eqref{eq:reflection_tm}. The free energy \eqref{eq:fr_dr}
and \eqref{eq:fr_die} calculated for $r$ round-trips thus scales like $f^{(r)} \sim \left(R_1 R_2/\mathcal{L}^2\right)^{3r}$ which means that the full expression is dominated by the single round-trip expression $r=1$. For two Drude spheres in vacuum, we obtain from the asymptotic expansion of \eqref{eq:f1_dr}
\begin{equation}
\label{eq:limdip_dr}
 f_{\mDvD,u\neq 0} \sim \frac{3}{8y^3}
\end{equation}
while we get from \eqref{eq:f1_dr_u0} for the special case of the plane-sphere geometry
\begin{equation}
\label{eq:limdip_dr0}
 f_{\mDvD,u=0} \sim \frac{1}{4y^3}\,.
\end{equation}
This leads to a ratio $2/3$ between the two asymptotic expressions at this limit. Note that in view of the definition \eqref{eq:def_y} of $y$, the reduced free energy decays with $L^{-6}$ for two spheres but only with $L^{-3}$ for the plane-sphere setup. 

For two dielectric spheres in an electrolyte, we get from \eqref{eq:f1_die}
\begin{equation}
\label{eq:limdip_die}
f_{\mded,u\neq 0} \sim \frac{3}{32y^3}
\end{equation}
and from \eqref{eq:f1_die_u0} for the plane-sphere geometry $u=0$
\begin{equation}
\label{eq:limdip_die0}
f_{\mded,u=0} \sim \frac{1}{8y^3}\,,
\end{equation}
with a ratio $4/3$ between the two results. We also note that
the Drude free energy is larger by a factor $2$ than that for dielectrics in the plane-sphere geometry ${(u=0)}$, whereas the ratio goes up to $4$ for the sphere-sphere geometry ($u\neq0$), with these results to be expected from the discussions devoted to the dipole limit in Ref. \onlinecite{Ingold2015}. 

In the short-distance limit, the reduced free energy $f^{(1)}$ does no longer depend on $u$ while the DvD and ded models tend to the same curve as the scalar model. In this limit however, we need to take multiple round-trips into account, which will be discussed in the next subsection.

\begin{figure}
\centering
\includegraphics[width=\columnwidth]{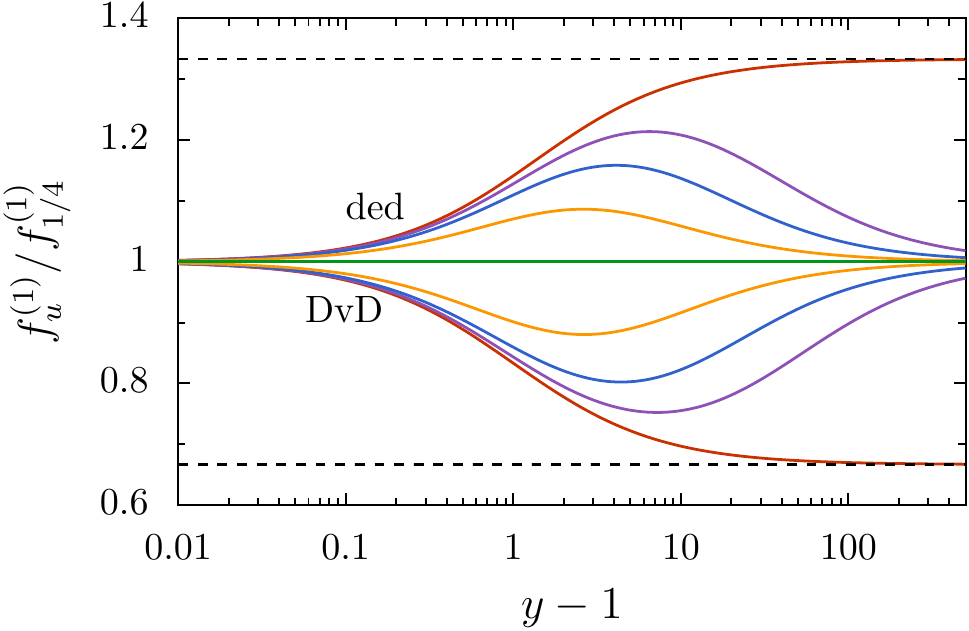}
\caption{Ratio $f_u^{(1)}/f_{1/4}^{(1)}$ for fixed values of $u$, with the same conventions as in Fig.~\ref{fig:FigFuvsYm}. All curves show variations very similar to those on Fig.~\ref{fig:ratioFuonF0vsYm}.}
\label{fig:FigFuonF0vsYm}
\end{figure}

The variations with $u$ of the curves corresponding to one round-trip are better analysed by showing in Fig.~\ref{fig:FigFuonF0vsYm} the ratio $f_u^{(1)}/f_{1/4}^{(1)}$ of the contribution calculated at $u$ by that at $u=0.25$. As previously, the red, violet, blue, yellow, and green curves correspond to $u=0, 0.016, 0.04, 0.1,$ and $0.25$. The ratios larger than 1 correspond to the ded model, and those smaller than 1 to the DvD model. For both models, the dependence on $u$ disappears at small-distances $y\to 1$. At large distances, the DvD and ded curves reach asymptotically the values 2/3 and 4/3 obtained in the dipolar limit if $u=0$. All curves show variations very similar to those on Fig.~\ref{fig:ratioFuonF0vsYm}, and this will be explained by results obtained below. 

\subsection{Ratio of the full expression to single round-trip contributions}

We go on in a quantitative understanding of the variation of the curves by looking now at the ratio of the full expression for $f_u$ to the one round-trip contribution ${f_u^{(1)}}$
\begin{equation}
\label{def_phi}
\phi_u = \frac{f_u}{f^{(1)}_u}\,.
\end{equation}  
This ratio goes to 1 at large distances as we know that the single round-trip expression dominates the full expression there. 

In the short-distance limit, we know that all curves tend to the same constant determined by the divergent part for $y\to1$ of the scalar case. This divergence is related to a zero eigenvalue of the matrix $\mathbf{M}_r^\sigma$ representing the bilinear form in the Gaussian integrands. As mentioned above, we can give this matrix an interpretation in terms of a tight-binding model. For the scalar case, $\mathbf{M}_r^\sigma$ describes a ring with periodic hopping matrix elements between neighboring sites where a zero eigenvalue may appear. The contributions for the DvD and ded models appearing in addition to the scalar contribution yield matrices corresponding either to a system with open boundary conditions or to generically disordered hopping matrix elements and thus cannot give rise to a zero eigenvalue. As a consequence their contribution cannot diverge for $y\to1$. We can thus deduce the divergent part of the reduced free energy from \eqref{eq:f1_sc} and \eqref{eq:fr_sc}.
Carrying out the sum over all round-trips according to \eqref{eq:def_f}, we obtain for the divergent
part in the short-distance limit of the reduced free energy
\begin{equation}
\label{eq:limpfa}
f \sim \frac{\zeta(3)}{8(y-1)}\,,
\end{equation}
where $\zeta(3)=\sum1/r^3=1.202\dots$ is Apéry's constant.
This result coincides with the proximity-force approximation, which gives the same expression for all models under study here. 

\begin{figure}
\centering
\includegraphics[width=\columnwidth]{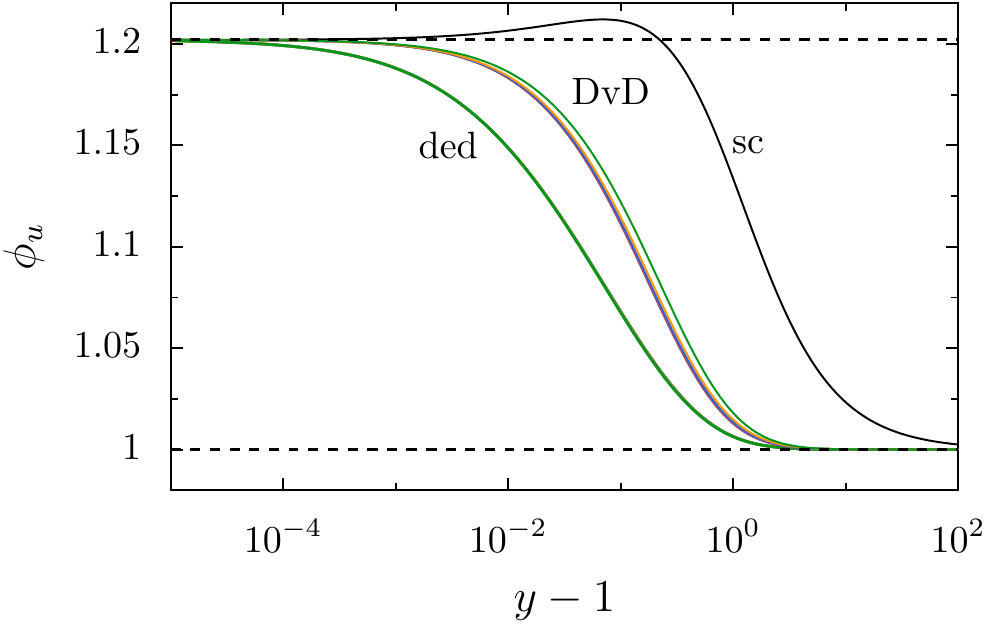}
\caption{Ratio $\phi_u = f_u/f^{(1)}_u$ drawn as functions of $y-1$ for 
fixed values of $u$ with the same conventions as on Fig. \ref{fig:FigFuvsYm}. The upper set of colored curves corresponds to the Drude case, the lower one to dielectric spheres in salted water. In both sets, the curves depend weakly on the parameter $u$ but they are not identical to each other. The black curve corresponds to the $u-$independent scalar case and is shown for comparison. }
\label{fig:FigPhiuvsYm}
\end{figure}

The functions $\phi_u$ defined in \eqref{def_phi} are depicted in Fig.~\ref{fig:FigPhiuvsYm} for the Drude spheres in vacuum and the dielectric spheres in an electrolyte. 
In both cases, the ratio is a monotonically decreasing function of $y-1$, at fixed $u$, which goes from the known value $\zeta(3)\simeq1.202$ at small distances to $1$ at large distances. 

For dielectric spheres in an electrolyte, it is remarkable that the ratio $\phi_u$ depends very weakly on the parameter $u$ with the curves drawn for different values of $u$ nearly indistinguishable.
The difference between the curves drawn for different $u$  
is more noticeable for Drude spheres in vacuum, but it remains small. 
The curve for the scalar case, $u-$independent, is also shown for comparison (black curve). It has the same limits as the two electromagnetic models, but is not varying monotonically, and it has quite different values from those met for the previous cases. Again the $u-$independence for the scalar model corresponds to exact conformal invariance while the weak dependence for electromagnetic models corresponds to a weak breaking of conformal invariance.

The reason for this weak dependence property can be understood qualitatively. The contributions of multiple round-trips are important when the single round-trip ones are themselves large, that is at short distances where all contributions tend to become the same.
A noticeable $u-$dependence might appear at large distances but $\phi_u$ is anyway close to unity there since $\kf_u$ and $\kf_u^{(1)}$ tend to become identical. 

\begin{figure}
\centering
\includegraphics[width=\columnwidth]{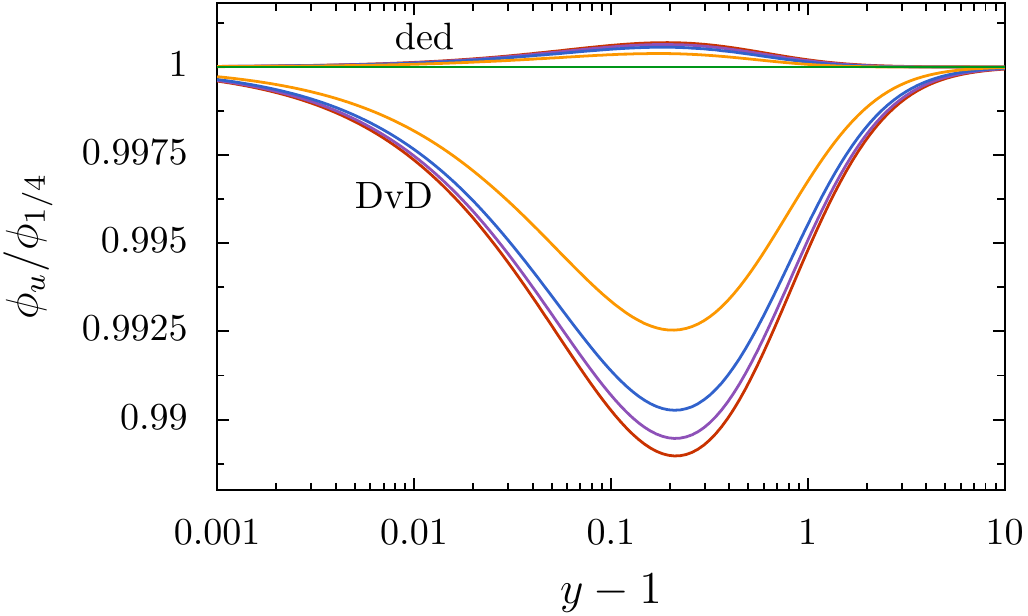}
\caption{Ratio $\phi_u/\phi_{1/4}$ drawn as functions of $y-1$ for fixed values of $u$ with the same conventions as on Fig. \ref{fig:FigFuvsYm}. The lower set of curves corresponds to the DvD case, the upper one to the ded model. The green curve corresponds to $u=1/4$ and also to the scalar case.}
\label{fig:FigPhiuonPhi0vsYm}
\end{figure}

The variation with $u$ of the $\phi-$curves for Drude and dielectric models can be assessed by plotting the ratio $\phi_u/\phi_{1/4}$ which is shown on Fig. \ref{fig:FigPhiuonPhi0vsYm}. Curves with ratio below 1 correspond to the DvD model, while those with ratio above 1 correspond to the ded model. The green curve stays at a ratio 1 as it represents the case $u=0.25$ as well as the result for the scalar case which is $u-$indepependent. Again, the small deviations of the ratio from 1 show a weak breaking of conformal invariance, in opposite directions for the DvD and ded models. 

The $\phi-$curves for Drude and dielectric models can be given a good approximation by a rational function $\phi_\mathrm{rm}$ of the argument $e^{y-1}$
\begin{equation}
\label{eq:epsilon}
\left|\frac{\phi_u(y)}{\phi_\mathrm{rm}(y)}-1
\right| < \epsilon\,.
\end{equation}
The rational model function $\phi_\mathrm{rm}$ is defined as in
Ref.~\onlinecite{Schoger2021universal}
\begin{equation}
\label{eq:rational_model}
\phi_\mathrm{rm} = \prod_{k=1}^n
\frac{e^{y-1}+ \nu_k -1}{e^{y-1}+ \mu_k -1}\,.
\end{equation}
The best model parameters $\nu_k,\, \mu_k$ have been determined 
for $n=2$ by finding the best fit $\phi_\mathrm{rm}$ on the whole domain of parameters (using a routine from SciPy \cite{Virtanen2020} for fitting the ratio $\phi_u(y)$). We determined the rational model for fixed $u= u_\mathrm{ref}$ and then calculated the maximal deviation $\epsilon$ as defined in eq.~\eqref{eq:epsilon}.
The corresponding model parameters calculated for $u_\mathrm{ref} = 0.1$ and $0.15$ respectively are given in Tab.~\ref{rm}. They lead to the very low value $\epsilon_{\mded}=1.2 \times 10^{-3}$ for the ded model and the low value $\epsilon_{\mDvD}=5.9 \times 10^{-3}$ for the DvD model.

\begin{table}[b]
\centering
\caption{Model parameters ($n=2$) to be used with \eqref{eq:rational_model} for the DvD model
(with $\epsilon_{\mDvD} = 5.9 \times 10^{-3}$)
and the ded model (with $\epsilon_{\mded} = 1.2 \times 10^{-3}$).}
{\begin{tabular}{c|cc||cc}
& \multicolumn{2}{c||}{DvD model }   
& \multicolumn{2}{c}{ded model}  \\
\hline
& $k=1$ & $k=2$ & $k=1$  & $k=2$ \\
\hline
$\nu_k$ &  0.011495 & 0.19868  &  0.004618  & 0.09639  \\
\hline
$\mu_k$ &  0.011359 & 0.16728  &  0.004415  & 0.08397 \\
\hline
\end{tabular}
\label{rm}}
\end{table}

We are left in the end of this reasoning with a good approximation of the full function $\kf_u(y)$ 
\begin{equation}
\kf_u(y) = \kf_u^{(1)}(y) \, \phi_\mfit(y)  ~.
\label{eq:approx}
\end{equation}
The first factor is the analytical one round-trip expression given in \eqref{eq:f1_dr} and \eqref{eq:f1_die} for the two models and depending on $y$ and $u$. Meanwhile the second factor $\phi_\mfit$ is the rational model \eqref{eq:rational_model} with the associated parameters given in Tab.~\ref{rm} for the two models. These approximations correspond to an accuracy $\epsilon$ which should be sufficient for most applications. Should a better accuracy be needed, a lower $\epsilon$ could be obtained with an higher order $n$ in the model \eqref{eq:rational_model}.

\section{Experimental evidence of the universal thermal interaction in the dielectric-electrolyte-dielectric configuration}

As mentioned in the introduction, the universal thermal
Casimir interaction corresponds to a more easily accessible domain of parameters in the ded configuration than in the previously known DvD one. 
Distances of the order of or larger than $\lambda_T\sim 8\,\mu$m are indeed required in the latter case, which lead to extremely weak signals. In contrast, the universal contribution is already dominant at much shorter distances, of the order of the characteristic length $\ell_T\sim 0.1 \,\mu$m, when considering typical dielectric materials interacting across electrolyte solutions. 

\begin{figure}
    \centering
    \includegraphics[width=\columnwidth]{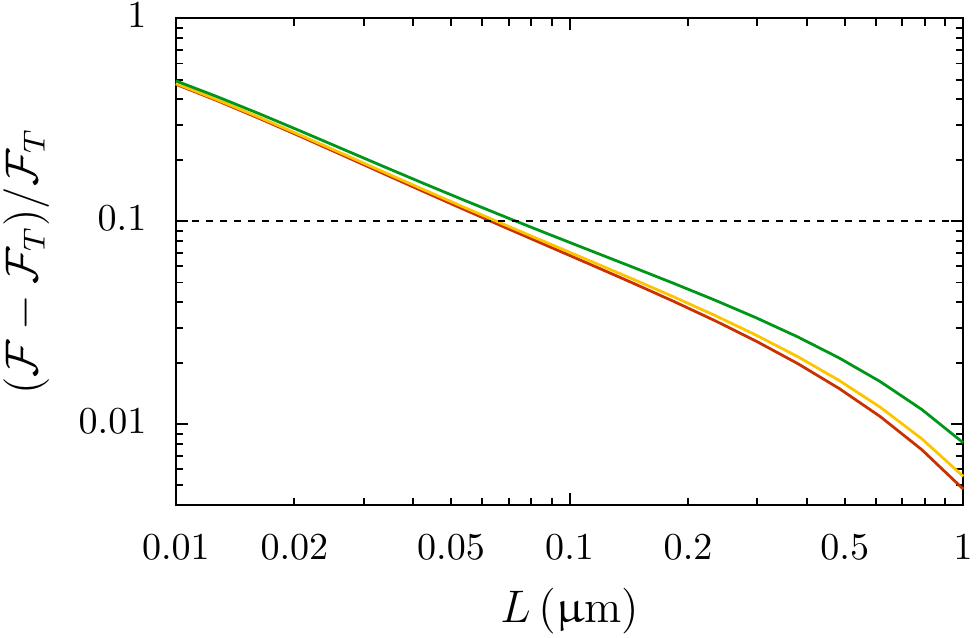}
    \caption{Ratio $(\mathcal{F}-\mathcal{F}_T)/\mathcal{F}_T$ with the full free energy $\mathcal{F}$ given in \eqref{eq:free_energy} and
    the high-temperature result $\mathcal{F}_T$ \eqref{eq:free_energy_ht} as function of the surface-to-surface distance $L$. The free energies were calculated for the setup of two silica spheres in water at $T=296\,$K as discussed in Ref.~\onlinecite{Pires2021} with $R_1 = 2.35\, \mu$m and ${u=0.14}$ (bright yellow). The dielectric functions of silica and water were modeled according to Ref.~\onlinecite{vanZwol2010}. The ratio was also calculated for $u=0$ (red) and $u=0.25$ (green) with the same $R_1$.}
    \label{fig:finite_frequency}
\end{figure}

The value of $\ell_T$ can be calculated more precisely, for example in the case of silica spheres, for which $\ell_T\approx 0.07\,\mu$m is estimated from the results shown in Fig.~\ref{fig:finite_frequency}. 
We consider the interaction between two silica spheres in salted water and plot the ratio $(\mathcal{F}-\mathcal{F}_T)/\mathcal{F}_T$ here as a function of shortest distance $L$ measured in $\mu$m, with a value of the smaller radius $R_1=2.35\,\mu$m and values of the larger one $R_2$ deduced from geometrical parameter $u=0$ (red), $0.14$ (yellow) and $0.25$ (green). We define $\ell_T$ as the distance at which the non-universal relative correction is $10\%$ of the universal thermal energy (dashed horizontal line on Fig.~\ref{fig:finite_frequency}). 

The universal thermal Casimir interaction between two silica microspheres in salted water was recently probed~\cite{Pires2021} by employing single-beam optical tweezers~\cite{Ashkin1986} in the distance range from $0.2$ to $0.5\,\mu$m. A silica microsphere of radius $R_1~=~2.35\,\mu$m was optically trapped while a second larger silica microsphere was attached to the glass coverslip at the bottom of the sample. 
The larger sphere had a radius $R_2~=~11.74\,\mu$m which corresponds to the geometric parameter $u\approx 0.14$.
The distance between the larger microsphere and the trapping laser beam was controlled by a piezoelectric nano-positioning stage. 

Optical tweezers were previously employed to probe colloidal interactions~\cite{Crocker1994,Hansen2005,Sainis2007,Elmahdy2010,Ether2015,Kundu2019} and to assess the non-additivity of the critical Casimir force~\cite{Paladugu2016}.
They are ideally suited for probing the universal Casimir interaction in the dielectric/electrolyte/dielectric configuration. Trapping conditions indeed favor dielectric materials such that the radiation pressure, associated to the reflection of the trapping laser beam, is dominated by the gradient force arising from refraction~\cite{Neto2000}.
Thus, optical tweezing is possible when the refractive indices of the 
dielectric material and of the electrolyte solution nearly match at the trapping laser wavelength, typically in the infrared.
In the context of the Casimir effect, the index matching reduces the contribution of nonzero Matsubara frequencies~\cite{Parsegian1971}, leading to rather low values of $\ell_T$ and favoring the universal thermal contribution in comparison with non-universal ones. In other words, single-beam trapping of particles is possible precisely when they are made of a dielectric material such that $\ell_T$ is conveniently small for probing the universal Casimir interaction.

In order to mimic typical conditions of biological interest, a NaI concentration of $0.22\,$M was employed in Ref.~\onlinecite{Pires2021}, corresponding to a Debye screening length $\lambda_\mD=0.64\,$nm. As a consequence, electrostatic interactions are completely suppressed~\cite{Israelachvili2011,Ether2018} at the distance range probed experimentally. In addition, the interaction resulting from thermal fluctuations of the electrostatic potential~\cite{Davies1972,Mitchell1974,MahantyNinham1976,Parsegian2006} is also screened across the distance $\lambda_{\rm D}$ and thus negligible.
Such effect can be interpreted as the contribution of longitudinal modes to the Casimir effect in a nonlocal medium~\cite{MaiaNeto2019}, and can be calculated exactly for the geometry of two spheres in salted water~\cite{Nunes2021}.

\begin{figure}
    \centering
    \includegraphics[width=\columnwidth]{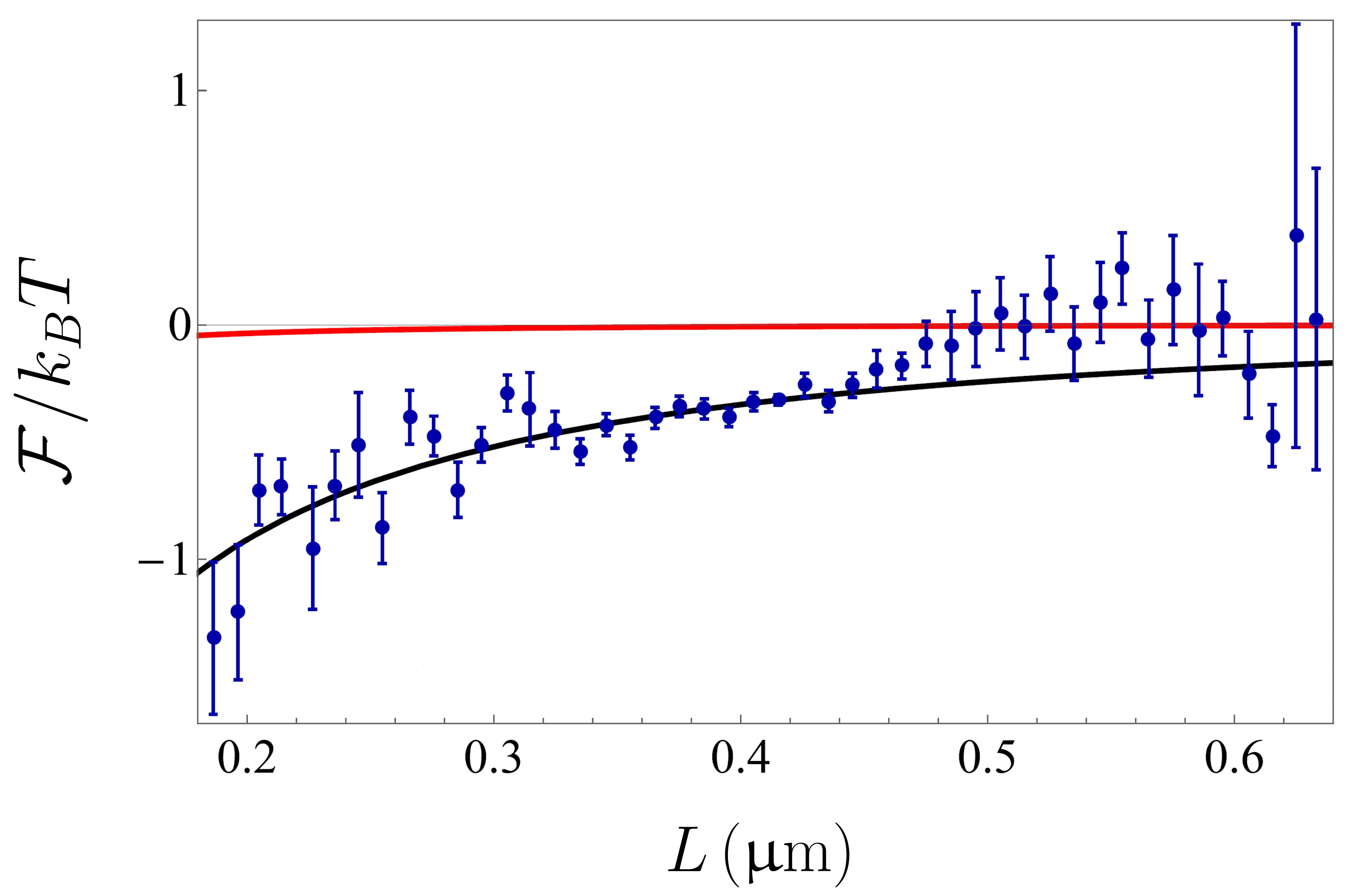}
    \caption{Interaction free energy (in units of $k_\mB T$)
    between two silica microspheres as 
    function of the surface-to-surface distance $L$: experimental points with errors bars, and theoretical Casimir energy including (black) or not (red) the universal thermal contribution. Adapted from Ref. \onlinecite{Pires2021} by Pires et al, used under CC BY.}
    \label{fig:theory_vs_experiment}
\end{figure}

The experimental results for the variation of the interaction energy 
with distance are shown in Fig.~\ref{fig:theory_vs_experiment}
as points with error bars. They are compared with two different theoretical curves. The red line represents the Casimir free energy calculated by excluding the universal thermal contribution. Clearly, this result fails to adequately describe the experimental data. Agreement is found only when including this contribution as represented by the theoretical black curve. 
Since the distances probed experimentally are such that $L > \ell_T,$ the Casimir interaction is completely dominated by the universal thermal effect. Indeed, the red curve shows that
the contribution of nonzero Matsubara frequencies is negligible in the probed range, as already discussed in connection with Fig.~\ref{fig:finite_frequency}. In short, the data shown in Fig.~\ref{fig:theory_vs_experiment} are direct evidence of the  universal Casimir interaction mediated by low-frequency unscreened
TM thermal fluctuations coupled to electric multipoles of each sphere. Such universal thermal effect, discussed in detail in the present paper, corresponds to the long-distance asymptotic behavior of the total Casimir interaction. For typical dielectric materials  interacting across salted water at room temperature, the universal contribution is already dominant at distances of the order of $\ell_T\sim 0.1\,\mu$m.

Note that a Casimir attraction stronger than the theoretical prediction excluding the universal contribution is also reported under similar experimental conditions in Ref.~\onlinecite{Hansen2005}, though it was not associated to a theoretical calculation.

\section{Conclusion}

In this paper, we have discussed analogous universality properties obtained for the high-temperature Casimir free energy between two spheres in two different physical situations. One corresponds to metallic spheres described by the Drude model in vacuum and the other to dielectric spheres in a conducting electrolyte. In both cases, universality comes from the fact that the static dielectric function of one of the two media is infinite while that of the other remains finite. In this sense, the two problems are dual of each other, with the Drude sphere and salted water exhibiting a finite static conductivity and thus an infinite dielectric constant. This explains in both cases why the results are independent of the details of the optical response of the involved materials. 

The difference between the two cases implies that the minimal distance over which the universal contribution dominates is not at all the same. In the case of vacuum, this distance is essentially the thermal wavelength $\simeq8\mu$m while a much smaller distance $\ell_T\sim 100\,$nm is found in the case of salted water. The universality property thus covers a much broader distance range in the latter case and this has made possible an experimental demonstration of the non-screened Casimir interaction between dielectric spheres in salted water with strong screening \cite{Pires2021}. 

This interaction should have important consequences for the physics of biological interfaces and of colloidal solutions. For these consequences to play an important role, the interaction has to represent a significant fraction of the thermal energy $k_\mB T$ as the immersion in a liquid medium imposes a Brownian motion to the spheres. We see on Fig.~\ref{fig:FigFuvsYm} that this condition is met when the distance $L$ is of the order or smaller than one tenth of the effective radius. In the associated domain $y-1~\lesssim~ 0.1$, the interaction is mainly determined by the conformally invariant parameter $y$ associated with the two-spheres geometry, with the dependence on $u$ remaining weak (see Fig.~\ref{fig:FigFuonF0vsYm}). In spite of these observations, the interaction cannot be deduced from any of the known limiting cases. A simple estimation can however be deduced from Eq.~\eqref{eq:approx} without having to carry out the full numerical calculation.

\section*{Acknowledgments}
The authors acknowledge stimulating discussions with G. R. Araujo, M. Borges, A. Canaguier-Durand, R. S. Decca, R. S. Dutra, D. S. Ether, S. Frases, R. Guérout, A. B. Moraes, M. J. B. Moura, H. M. Nussenzveig, L. B. Pires, B. Pontes, F. S. S. Rosa, N. B. Viana. 
P.A.M.N. thanks Sorbonne Université for hospitality and
acknowledges funding from the Brazilian agencies Conselho Nacional de Desenvolvimento Cient\'{\i}fico e Tecnol\'ogico (CNPq--Brazil),
Coordenaç\~ao de Aperfeiçamento de Pessoal de N\'{\i}vel Superior (CAPES--Brazil),  Instituto Nacional de Ci\^encia e Tecnologia Fluidos Complexos  (INCT-FCx), and the
Research Foundations of the States of Minas Gerais (FAPEMIG), Rio de Janeiro (FAPERJ) and S\~ao Paulo (FAPESP).


\begin{thebibliography}{99}
\expandafter\ifx\csname urlstyle\endcsname\relax
  \providecommand{\doi}[1]{doi:\discretionary{}{}{}#1}\else
  \providecommand{\doi}{doi:\discretionary{}{}{}\begingroup
  \urlstyle{rm}\Url}\fi

\bibitem{Casimir1948}
H.~B.~G. Casimir, {\em {Proc. Kon. Ned. Akad. Wetenschappen}} {\bf 51},  79
  ({1948}).

\bibitem{Lamoreaux1999}
S.~K. Lamoreaux, {\em American Journal of Physics} {\bf 67}, 850  (1999),
  \doi{10.1119/1.19138}.

\bibitem{Milton2011}
K.~A. Milton, {\em American Journal of Physics} {\bf 79}, 697  (2011),
  \doi{10.1119/1.3573976}.

\bibitem{Decca2011}
R.~Decca, V.~Aksyuk and D.~L\'opez, Casimir force in micro and nano electro
  mechanical systems, in {\em Casimir Physics\/},  eds. D.~Dalvit, P.~Milonni,
  D.~Roberts and F.~da~Rosa, Lecture Notes in Physics Vol.~834 (Springer,
  2011), ch.~9, p. 287.

\bibitem{Gong2021}
T.~Gong, M.~R. Corrado, A.~R. Mahbub, C.~Shelden and J.~N. Munday, {\em
  Nanophotonics} {\bf 10},   523  (2021), \doi{10.1515/nanoph-2020-0425}.

\bibitem{CanaguierDurand2012}
A.~Canaguier-Durand, G.-L. Ingold, M.-T. Jaekel, A.~Lambrecht, P.~A. Maia~Neto
  and S.~Reynaud, {\em {Phys. Rev. A}} {\bf {85}},   {052501}  ({2012}),
  \doi{10.1103/PhysRevA.85.052501}.

\bibitem{Bimonte2012}
G.~Bimonte and T.~Emig, {\em {Phys. Rev. Letters}} {\bf {109}},   {160403}
  ({2012}), \doi{10.1103/PhysRevLett.109.160403}.

\bibitem{Bostrom2000}
M.~Bostr\"om and B.~E. Sernelius, {\em Phys. Rev. Letters} {\bf 84}, 4757
  (2000), \doi{10.1103/PhysRevLett.84.4757}.

\bibitem{Sushkov2011}
A.~O. Sushkov, W.~J. Kim, D.~A.~R. Dalvit and S.~K. Lamoreaux, {\em Nature
  Physics} {\bf 7}, 230  (2011), \doi{10.1038/nphys1909}.

\bibitem{MaiaNeto2019}
P.~A. Maia~Neto, F.~S.~S. Rosa, L.~B. Pires, A.~B. Moraes, A.~Canaguier-Durand,
  R.~Guérout, A.~Lambrecht and S.~Reynaud, {\em {European Physical Journal D}}
  {\bf 73},   {178}  ({2019}), \doi{10.1140/epjd/e2019-100225-8}.

\bibitem{Schoger2021universal}
T.~Schoger, B.~Spreng, G.-L. Ingold, P.~A. Maia~Neto and S.~Reynaud, {
  Universal {Casimir} interaction between two dielectric spheres in salted
  water} https://arxiv.org/abs/2112.08800,  (2021).

\bibitem{Pires2021}
L.~B. Pires, D.~S. Ether, B.~Spreng, G.~R.~S. Araujo, R.~S. Decca, R.~S. Dutra,
  M.~Borges, F.~S.~S. Rosa, G.-L. Ingold, M.~J.~B. Moura, S.~Frases, B.~Pontes,
  H.~M. Nussenzveig, S.~Reynaud, N.~B. Viana and P.~A. Maia~Neto, {\em {Phys.
  Rev. Research}} {\bf {3}},   {033037}  ({2021}),
  \doi{10.1103/PhysRevResearch.3.033037}.

\bibitem{Lambrecht2006}
A.~Lambrecht, P.~A. Maia~Neto and S.~Reynaud, {\em {New Journal of Physics}}
  {\bf {8}},   {243}  ({2006}), \doi{10.1088/1367-2630/8/10/243}.

\bibitem{MaiaNeto2008}
P.~A. Maia~Neto, A.~Lambrecht and S.~Reynaud, {\em Phys. Rev. A} {\bf 78},
  012115  (2008), \doi{10.1103/PhysRevA.78.012115}.

\bibitem{Fisher1978}
M.~E. Fisher and P.-G. de~Gennes, {\em {Comptes Rendus Acad\'emie des Sciences
  Paris B}} {\bf 287},   207–209  ({1978}).

\bibitem{Hanke1998}
A.~Hanke, F.~Schlesener, E.~Eisenriegler and S.~Dietrich, {\em Phys. Rev.
  Letters} {\bf 81},   1885  (1998), \doi{10.1103/PhysRevLett.81.1885}.

\bibitem{Hertlein2008}
C.~Hertlein, L.~Helden, A.~Gambassi, S.~Dietrich and C.~Bechinger, {\em Nature}
  {\bf 451}, 172  (2008), \doi{10.1038/nature06443}.

\bibitem{Magazzu2019}
A.~Magazzu, A.~Callegari, J.~Pablo~Staforelli, A.~Gambassi, S.~Dietrich and
  G.~Volpe, {\em {Soft Matter}} {\bf {15}},   {2152}  ({2019}),
  \doi{10.1039/c8sm01376d}.

\bibitem{Parsegian2006}
V.~A. Parsegian, {\em Van der Waals forces} (Cambridge University Press, 2006).

\bibitem{Gambassi2009}
A.~Gambassi, {\em Journal of Physics: Conference Series} {\bf 161},   012037
  (2009), \doi{10.1088/1742-6596/161/1/012037}.

\bibitem{RodriguezLopez2011}
P.~Rodriguez-Lopez, {\em Phys. Rev. B} {\bf 84},   075431  (2011),
  \doi{10.1103/PhysRevB.84.075431}.

\bibitem{Teo2012}
L.~P. Teo, {\em Phys. Rev. D} {\bf 85},   045027  (2012),
  \doi{10.1103/PhysRevD.85.045027}.

\bibitem{Bimonte2018a}
G.~Bimonte, {\em Phys. Rev. D} {\bf 97},   085011  (2018),
  \doi{10.1103/PhysRevD.97.085011}.

\bibitem{Bimonte2018b}
G.~Bimonte, {\em Phys. Rev. D} {\bf 98},   105004  (2018),
  \doi{10.1103/PhysRevD.98.105004}.

\bibitem{CanaguierDurand2010}
A.~Canaguier-Durand, P.~A. Maia~Neto, A.~Lambrecht and S.~Reynaud, {\em Phys.
  Rev. Letters} {\bf 104},   040403  (2010),
  \doi{10.1103/PhysRevLett.104.040403}.

\bibitem{Schwinger1978}
J.~Schwinger, L.~L. DeRaad and K.~A. Milton, {\em {Annals of Physics}} {\bf
  115}, 1  ({1978}), \doi{10.1016/0003-4916(78)90172-0}.

\bibitem{Guerout2014}
R.~Guérout, A.~Lambrecht, K.~A. Milton and S.~Reynaud, {\em Phys. Rev. E} {\bf
  90},   042125  (2014), \doi{10.1103/PhysRevE.90.042125}.

\bibitem{Spreng2020}
B.~Spreng, P.~A. Maia~Neto and G.-L. Ingold, {\em {Journal of Chemical
  Physics}} {\bf {153}},   024115  ({2020}), \doi{10.1063/5.0011368}.

\bibitem{Schoger2021}
T.~Schoger and G.-L. Ingold, {\em {SciPost Phys. Core}} {\bf 4},  11
  ({2021}), \doi{10.21468/SciPostPhysCore.4.2.011}.

\bibitem{Spreng2018}
B.~Spreng, M.~Hartmann, V.~Henning, P.~A. Maia~Neto and G.-L. Ingold, {\em
  {Phys. Rev. A}} {\bf {97}},   062504  ({2018}),
  \doi{10.1103/PhysRevA.97.062504}.

\bibitem{BohrenHuffman2004}
C.~F. Bohren and D.~R. Huffman, {\em Absorption and {Scattering} of {Light} by
  {Small} {Particles}} (Wiley-VCH, 2004),  Chap. 4.

\bibitem{Eisenriegler1995}
E.~Eisenriegler and U.~Ritschel, {\em Phys. Rev. B} {\bf 51}, 13717  (1995),
  \doi{10.1103/PhysRevB.51.13717}.

\bibitem{Zhao2013}
R.~Zhao, Y.~Luo, A.~I. Fernández-Domínguez and J.~B. Pendry, {\em Phys. Rev.
  Letters} {\bf 111},   033602  (2013), \doi{10.1103/PhysRevLett.111.033602}.

\bibitem{Teo2014}
L.~P. Teo, {\em Phys. Rev. D} {\bf 90},   045012  (2014),
  \doi{10.1103/PhysRevD.90.045012}.

\bibitem{Molinari1997}
L.~Molinari, {\em J. Phys. A} {\bf 30}, 983  (1997),
  \doi{10.1088/0305-4470/30/3/021}.

\bibitem{Molinari2008}
L.~G. Molinari, {\em Linear Algebra Appl.} {\bf 429}, 2221  (2008),
  \doi{10.1016/j.laa.2008.06.015}.

\bibitem{Burkhardt1995}
T.~W. Burkhardt and E.~Eisenriegler, {\em Phys. Rev. Letters} {\bf 74}, 3189
  (1995), \doi{10.1103/PhysRevLett.74.3189}.

\bibitem{Thomson1847}
W.~Thomson, {\em J. Math. Pures Appl. 1e Sér.} {\bf
  {12}}, 256  (1847).

\bibitem{Thomson1853}
W.~Thomson, {\em The London, Edinburgh, and Dublin Philosophical Magazine and
  Journal of Science} {\bf {5}},  287  ({1853}),
  \doi{10.1080/14786445308647247}.

\bibitem{Maxwell1873}
J.~C. Maxwell, {\em {A Treatise on Electricity and Magnetism}} ({Oxford
  University Press, 1873, reprinted by Dover Publications}, {1954}).

\bibitem{Liouville1847}
J.~Liouville, {\em Journal de mathématiques pures et appliquées 1e série}
  {\bf {12}}, 265  (1847).

\bibitem{Bromwich1900}
T.~J.~I. Bromwich, {\em Proceedings of the London Mathematical Society} {\bf
  s1-33}, 185  (1900), \doi{10.1112/plms/s1-33.1.185}.

\bibitem{Klein1926}
F.~Klein, Lehre von den Transformationen, in {\em Vorlesungen {\"u}ber
  H{\"o}here Geometrie\/},  ed. W.~Blaschke (Springer Berlin Heidelberg, 1926),
  pp. 136--309.

\bibitem{Fosco2016}
C.~D. Fosco, F.~C. Lombardo and F.~D. Mazzitelli, {\em Phys. Rev. D} {\bf 93},
   125015  (2016), \doi{10.1103/PhysRevD.93.125015}.

\bibitem{Lekner2011}
J.~Lekner, {\em Journal of Electrostatics} {\bf 69}, 11  (2011),
  \doi{10.1016/j.elstat.2010.10.002}.

\bibitem{Smythe1988}
W.~Smythe, {\em Static and dynamic electricity}, 3rd edn. (Hemisphere
  Publishing, 1988).

\bibitem{Ingold2015}
G.-L. Ingold, S.~Umrath, M.~Hartmann, R.~Gu\'erout, A.~Lambrecht, S.~Reynaud
  and K.~A. Milton, {\em Phys. Rev. E} {\bf 91},   033203  (2015),
  \doi{10.1103/PhysRevE.91.033203}.

\bibitem{Virtanen2020}
{P. Virtanen et al.}, {\em Nature Methods} {\bf 17}, 261   (2020),
  \doi{10.1038/s41592-019-0686-2}.

\bibitem{Ashkin1986}
A.~Ashkin, J.~M. Dziedzic, J.~E. Bjorkholm and S.~Chu, {\em Opt. Letters} {\bf
  11},   288  (1986), \doi{10.1364/OL.11.000288}.

\bibitem{vanZwol2010}
P.~J. van Zwol and G.~Palasantzas, {\em Phys. Rev. A} {\bf 81},   062502
  (2010), \doi{10.1103/PhysRevA.81.062502}.

\bibitem{Crocker1994}
J.~C. Crocker and D.~G. Grier, {\em Phys. Rev. Letters} {\bf 73},   352
  (1994), \doi{10.1103/PhysRevLett.73.352}.

\bibitem{Hansen2005}
P.~M. Hansen, J.~K. Dreyer, J.~Ferkinghoff-Borg and L.~Oddershede, {\em Journal
  of Colloid and Interface Science} {\bf 287},   561  (2005),
  \doi{10.1016/j.jcis.2005.01.098}.

\bibitem{Sainis2007}
S.~K. Sainis, V.~Germain and E.~R. Dufresne, {\em Phys. Rev. Letters} {\bf 99},
    018303  (2007), \doi{10.1103/PhysRevLett.99.018303}.

\bibitem{Elmahdy2010}
M.~M. Elmahdy, C.~Gutsche and F.~Kremer, {\em The Journal of Physical Chemistry
  C} {\bf 114},   19452  (2010), \doi{10.1021/jp107673f}.

\bibitem{Ether2015}
D.~S. Ether, Jr., L.~B. Pires, S.~Umrath, D.~Martinez, Y.~Ayala, B.~Pontes,
  G.~R. d.~S. Araujo, S.~Frases, G.-L. Ingold, F.~S.~S. Rosa, N.~B. Viana,
  H.~M. Nussenzveig and P.~A. Maia~Neto, {\em {EPL}} {\bf {112}},   {44001}
  ({2015}), \doi{10.1209/0295-5075/112/44001}.

\bibitem{Kundu2019}
A.~Kundu, S.~Paul, S.~Banerjee and A.~Banerjee, {\em Applied Physics Letters}
  {\bf 115},   123701  (2019), \doi{10.1063/1.5110581}.

\bibitem{Paladugu2016}
S.~Paladugu, A.~Callegari, Y.~Tuna, L.~Barth, S.~Dietrich, A.~Gambassi and
  G.~Volpe, {\em Nature communications} {\bf 7},  1  (2016),
  \doi{10.1038/ncomms11403}.

\bibitem{Neto2000}
P.~A. Maia~Neto and H.~M. Nussenzveig, {\em Europhys. Letters} {\bf 50},   702
  (2000), \doi{10.1209/epl/i2000-00327-4}.

\bibitem{Parsegian1971}
V.~Parsegian and B.~Ninham, {\em Journal of Colloid and Interface Science} {\bf
  37}, 332  (1971), \doi{10.1016/0021-9797(71)90301-8}.

\bibitem{Israelachvili2011}
J.~N. Israelachvili, {\em {Intermolecular and Surface Forces, 3rd ed.}}
  ({Academic Press}, {2011}).

\bibitem{Ether2018}
D.~S. Ether, F.~S.~S. Rosa, D.~M. Tibaduiza, L.~B. Pires, R.~S. Decca and P.~A.
  Maia~Neto, {\em {Phys. Rev. E}} {\bf {97}},   {022611}  ({2018}),
  \doi{10.1103/PhysRevE.97.022611}.

\bibitem{Davies1972}
B.~Davies and B.~W. Ninham, {\em The Journal of Chemical Physics} {\bf 56},
  5797  (1972), \doi{10.1063/1.1677118}.

\bibitem{Mitchell1974}
D.~J. Mitchell and P.~Richmond, {\em J. Colloid Interface Sci.} {\bf 46},   118
   (1974), \doi{10.1016/0021-9797(74)90031-9}.

\bibitem{MahantyNinham1976}
J.~Mahanty and B.~W. Ninham, {\em Dispersion forces} (Academic Press, 1976).

\bibitem{Nunes2021}
R.~O. Nunes, B.~Spreng, R.~de~Melo~e Souza, G.-L. Ingold, P.~A. Maia~Neto and
  F.~S.~S. Rosa, {\em {Universe}} {\bf {7}},   {156}  ({2021}),
  \doi{10.3390/universe7050156}.

\end{thebibliography}

\end{document}